\documentclass[aps, pra,floats,tightenlines, a4paper,showpacs,twocolumn,10pt,superscriptaddress]{revtex4-1}
\usepackage{color,graphicx,bbm, amsmath, amssymb, amsthm, bm,textcomp,geometry,dsfont}
\usepackage{pbox,hyperref,array}
\usepackage{ulem}

\newcommand{\ket}[1]{\left|#1\right\rangle}

\newcommand{\bra}[1]{\left\langle #1\right|}

\newcommand{\D}{\text{d}}

\newcommand{\be}{\begin{equation}}
\newcommand{\ee}{\end{equation}}
\newcommand{\bea}{\begin{eqnarray}}
\newcommand{\eea}{\end{eqnarray}}

\newcommand{\fig}[1]{Fig.~\ref{#1}}
\newcommand{\eq}[1]{Eq.~(\ref{#1})}
\newcommand{\Sec}[1]{Sec.~\ref{#1}}

\newcommand{\I}{\ensuremath{{\mkern1mu\mathrm{i}\mkern1mu}}}

\begin{document}

\title{Noisy distributed sensing in the Bayesian regime}
\author{S. W\"olk}
 \affiliation{Institut f\"ur Theoretische Physik, Universit\"at Innsbruck, Technikerstra{\ss}e 21a, 6020 Innsbruck, Austria}
 \affiliation{Deutsches Zentrum f\"ur Luft und Raumfahrt e.V. (DLR), Institut f\"ur Quantentechnologie, S\"oflingerstr. 100, 89077 Ulm, Germany}
\author{P. Sekatski}
\affiliation{Departement Physik, Universit\"at Basel, Klingelbergstra{\ss}e 82, 4056 Basel, Switzerland}
 \author{ W.~D\"ur}
  \affiliation{Institut f\"ur Theoretische Physik, Universit\"at Innsbruck, Technikerstra{\ss}e 21a, 6020 Innsbruck, Austria}\

\date{\today}

\begin{abstract}
We consider non-local sensing of scalar signals with specific spatial dependence in the Bayesian regime. We design schemes that allow one to achieve optimal scaling and are immune to noise sources with a different spatial dependence than the signal. This is achieved by using a sensor array of spatially separated sensors and constructing a multi-dimensional decoherence free subspace. While in the Fisher regime with sharp prior and multiple measurements only the spectral range $\Delta$ is important, in single-shot sensing with broad prior the number of available energy levels $L$ is crucial. We study the influence of $L$ and $\Delta$ also in intermediate scenarios, and show that these quantities can be optimized separately in our setting. This provides us with a flexible scheme that can be adapted to different situations, and is by construction insensitive to given noise sources.

\end{abstract}

\maketitle

\section{Introduction}
Quantum sensing or quantum metrology \cite{Giovannetti_2011,Toth_2014,Pezze_2018} is one of the most promising applications of an upcoming quantum technology. Measuring quantities with ever higher precision lies at the heart of most natural sciences, and accordingly high precision measurements are a tool of uttermost importance. Quantum devices offer in principle a quadratic scaling advantage in the number of sensors, and have hence been studied in detail in recent years. Whenever some unknown signal or function should be sensed using multiple sensors, one is typically faced with a situation that the sensors are at different positions. This is the case for trapped ions \cite{Keller2019,Schindler2013, Blatt2008} as well as for arrays of superconducting qubits,  quantum dots or nitrogen-vacancy centers \cite{Childress2013, Kessler_2014}. Furthermore, with the rapid developments in quantum networks, even arrays of such quantum sensors distributed over large distances are into reach. Since a single quantum system or qubit is already a quantum sensor, any such arrangement of multiple qubits corresponds to a quantum sensor network \cite{Kimble2008, Blatt2008, Schindler2013, Childress2013,Kessler_2014, Wehner2018}. These networks can be used to measure non-local properties such as field gradients or spatial Fourier coefficients \cite{Urizar2013, Altenburg2017, Apellaniz2017,Sekatski2019}, or to increase the precision of atomic clocks, interferometers and telescope networks \cite{Komar2014, Komar2016, Landini2014, Ciampini2016, Khabiboulline2019, Khabiboulline2019b}. While the spatial distribution of sensors is irrelevant for signals without spatial dependence (as often considered in metrological scenarios), this is a crucial asset in the sensing of signals with certain spatial correlations.

In this paper we study the sensing of scalar, spatial dependent signals and show that one can indeed make use of such spatial correlations. By choosing appropriate quantum states of the sensors, one can make the sensor array sensitive only to a particular signal with a specific spatial dependence. This allows one to lock in to any signal of choice and measure only this signal. 
In this way, one can construct decoherence free subspaces for arbitrary given noise sources  \cite{Sekatski2017PRX, Sekatski2017quantummetrology,Dur_2014,Arrad_2014,Kessler_2014,Sekatski_2016,Zhou2018,Altenburg2016,Landini2014} and overcome the known vulnerability of metrological schemes under noise \cite{Fujiwara2008,Escher2011,Escher2012,Kolodynski2012, Sekatski2017PRX,Sekatski2017quantummetrology}. Such a decoherence free subspace (DFS) in quantum computation or standard quantum metrology is typically thought of to be available only in very specific situations, mainly when there is some correlated or restricted kind of noise. For noise source with a known spatial dependence one can essentially always construct such a DFS. The only requirement is that the spatial dependences of the signal to be sensed, and the one of the noise source are different. In any such case, one can find sensor states that are insensitive to a single or even multiple noise sources, while still being sensitive to the signal \cite{Sekatski2019}. In fact, a sensor array of $N+1$ sensors allows one to be insensitive to $N$ noise sources with different spatial dependences, and sense one specific signal. Notice that this insensitivity only refers to noise sources with different spatial dependence, but clearly a fluctuating constant noise field still jeopardizes the sensing of an constant signal field. However, e.g. in situations with different sources and decaying field strength with certain distance dependence $r^{-\alpha}$ --which is a rather typical situation an many physical set-ups--, these fields are linearly independent whenever they are located at different positions and sampled on fixed sensor positions using such a sensor network. Hence a DFS and lock-in to a specific signal can be constructed. This holds true under generic conditions, and the existence of a DFS is thus typical and not exceptional.

In \cite{Sekatski2019} such a sensing scheme was introduced and analyzed in the so-called Fisher regime, where the parameter $\varphi$ to be sensed is already approximately know, and multiple repetitions of the same experiment are considered. In this case the optimal state for sensing is given by a GHZ-type state, in the noiseless case just a superposition of two eigenstates of the signal Hamiltonian or generator $\hat{G}$ with minimal and maximal eigenvalue respectively. The achievable accuracy is given by the quantum Fisher information (QFI) \cite{Braunstein1994, Giovannetti2006, Toth2014}. In a scenario with multiple noise sources, a two-dimensional DFS that contains two states with different eigenvalues and a certain spectral distance $\Delta$ can be generically constructed, and hence the above mentioned features can be achieved. This ensures that Heisenberg scaling, i.e. a quadratic enhancement over the best classical protocol, can be obtained even in the presence of additional noise sources.
The situation is different in the so called Bayesian regime where the unknown parameter is specified by a (broad) probability distribution and only a single or few measurements can be performed. In this case, it is known that multiple states with different eigenvalues w.r.t. the signal Hamiltonian are required to achieve Heisenberg scaling\cite{Berry2000,Chiribella2004,Sekatski2017singlequbit}. Therefore, not only the spectral range $\Delta$ but also the number of different eigenvalues $L$ covered by a probe state are important for quantum metrology in the Bayesian regime, as we will further investigate in this paper.

In quantum sensor networks, it is often possible to maximize either the spectral range or the number of levels of a generator, by e.g. placing the sensors at appropriate positions, but not both simultaneously. Thus, it is important to know how $\Delta$ and $L$ will influence the precision for different scenarios as we will discuss in this paper. Moreover, not the total number $L$ is important in noisy scenarios but the effective number of levels within the decoherence-free subspace. For a single-shot scenario, i.e. when considering only a single run of preparing a probe state, letting it evolve and then measure the resulting state, the number of available levels $L$ is the crucial quantity as long as the measurement time can be freely chosen and is not considered to be a resource. The longer the evolution time, the larger the required number of levels. However, the spectral range $\Delta$ enters in the required evolution time,  as the strength of the signal is proportional to $\Delta$. Hence of a fixed time, both $\Delta$ and $L$ are important.  In this paper, we will introduce different methods to create a large number of linear spaced levels within the decoherence-free subspace. We introduce different methods with a different trade-off between increasing the number of effective levels $L$ within the decoherence-free subspace and the maximal achievable spectral range $\Delta$. Depending on the exact situations, as discussed in the first part of this paper, we can than choose a corresponding method to either maximally increase $L$, $\Delta$, or to share the provided resources to increase both simultaneously.

The main results of this paper can be summarized as follows:
(i) We analyze the effect of number of levels $L$ and spectral range $\Delta$ for Bayesian metrology with flat prior.
(ii) We provide a general way to construct multi-dimensional decoherence free subspaces with quantum sensor networks for spatially correlated scalar signals.
(iii) We show how to measure specific signals with a particular spatial dependence and a given prior, being completely insensitive to noise sources with a different spatial dependence.

The paper is organized as follows: First, we introduce the setup and summarize our results from \cite{Sekatski2019} in \Sec{sec:background}. Then, we start our investigation by  discussing different measurement scenarios in the Bayesian regime and the influence of $\Delta$ and $L$ on the precision in \Sec{sec:L_vs_D}. Consecutively, we described methods to create effective linear spectra within the decoherence free subspace by either increasing the internal degree of freedom of the sensors (\Sec{sec:spectra}) or by changing the position of the different sensors (\Sec{sec:position}). At the end, we summarize our results in \Sec{sec:conclusion} by comparing the different situations and methods.


\section{Setting and background\label{sec:background}}

In the following, we investigate methods to achieve maximal precision for estimating the unknown field strength $\omega$ of a global field $B_0(r)=\omega f_0(r)$ with given spatial distribution $f_0(r)$. For this purpose, we consider quantum sensor networks with $J$ sensors located at positions $r_j$. The time evolution of each local sensor is described by the local operator $\hat{Z}_j$ equal to the sum of all Pauli-z matrices of qubits located at $r_j$. The unknown phase $\varphi_0=\omega t$ is generated by the global generator
\be
\hat{G}_0 = \sum\limits_{j=1}^J f_0(r_j)\hat{Z}_j \label{eq:G}
\ee
via the time evolution $U=\exp(-it \omega \hat{G_0})$. Throughout this paper, we investigate situations where additional noise sources are present. These noise sources are describe via similar generators $\hat{G}_k$ with $1\leq k\leq K$ but with different spatial distributions $f_k(r_j)$. Strictly speaking, we assume that the vectors $\mathbf{f}_k=(f_k(r_1),\cdots, f_k(r_J))$ are linear independent. The state $\rho$ of the quantum sensor network after evolving for a time $t$ is given by
\be
\int \exp\left(-\I\sum\limits_{k=0}^K \varphi_k\hat{G}_k\right)\rho \exp\left(\I\sum\limits_{k=0}^K \varphi_k\hat{G}_k\right) \D\varphi_1 \cdots \D\varphi_K.
\ee
As a consequence, the coherence between two spin eigenstates $\mathbf{s}=(s_1,\cdots, s_J)$ with $\hat{Z}_j\ket{s_j}=s_j$ is destroyed whenever there exists at least one $k>0$ with 
\be
 \mathbf{f}_k(\mathbf{s}-\mathbf{s}')\neq 0
\ee
preventing us from obtaining information about the unknown phase $\varphi_0$.
Thus, optimal probe state consists of a superposition of eigenstates $\mathbf{s}$ which are all orthogonal to $\lbrace \mathbf{f}_k \rbrace$ for $1\leq k\leq K$ as we have demonstrated in \cite{Sekatski2019}. A priori the components $s_j$ can only take on integer multiples of $1/2$ which prevents us from creating spin vectors $\mathbf{s}$ orthogonal to $\lbrace \mathbf{f}_k\rbrace$ in certain cases. However, we can circumvent this restriction by adding dynamical control. Here, all spins at a corresponding site are switched at an intermediate time $t_j$ leading to effective spin components $s_j$ equal to non-integer multiples of $1/2$. In general, such orthogonal spin vectors can be created whenever there exists more probes $J>K$ than noises sources $K$. 

Optimal probe states in the Fisher regime (narrow prior, many measurements) consists of the superposition of the two effective spin eigenstates $\ket{\pm \mathbf{s}}$ which maximize the absolute value of the scalar product $\mathbf{s}\mathbf{f}_\perp$. Here, $\mathbf{f}_\perp$ denotes the component of $\mathbf{f}_0$ which is orthogonal to $\lbrace \mathbf{f}_k\rbrace$ with $1\leq k\leq K$. However,  the total number of distinct spin eigenstates on which the probe state is supported sets a limit on the amount of information on the field strength that can be gathered by the probe in a single run \cite{holevo1973}. Thus in the Bayesian regime also effective intermediate levels with 
$|\mathbf{s}\mathbf{f}_\perp|<\text{max}|\mathbf{s}\mathbf{f}_\perp|$ play an important role as we will discuss in the next sections.

\section{Spectral range versus number of levels\label{sec:L_vs_D}}

Let us first concentrate on achievable precisions for parameter estimation in the Bayesian regime without noise. In general, the precision of the estimation of $\omega$ depends on the spectral range $\Delta=\Gamma_\text{max}-\Gamma_\text{min}$, given by the difference of the maximal and minimal eigenvalue $\Gamma$ of $\hat{G}_0$,  and the number of levels $L$ of the generator $\hat{G}_0$.  Both of them depend on the spatial distribution of the sensors. Thus, generators with linear spectrum and different spectral range and different number of levels can be achieved by rearranging the local sensors. Usually, $\Delta$ and $L$ cannot be maximized simultaneously. As a consequence, it is important to know how the precision scales with $\Delta$ and $L$. Therefore, we will investigate this scaling for a couple of exemplary scenarios in this sections before we investigate achievable effective $\Delta$ and $L$ for noisy distributed sensing  in the next section.

\subsection{Single-shot estimation }

We will start our investigations in the Bayesian regime where we assume flat priors and single-shot estimation. The extremal case of a flat prior is an equally distributed unkown phase $\varphi=\omega t$ with prior $p(\varphi)=1/(2\pi)$ for $0\leq \varphi<2\pi$. For this situation,  Berry and Wiseman \cite{Berry2000} determined the optimal probe state for a generator
\be
\hat{G}_\text{BW}=\sum\limits_{j=1}^N \hat{z}_j.
\ee
with linear spectrum. Here, $\hat{z}_k$ denotes the Pauli-z matrix of a single qubit. Berry and Wiseman proved that the phase $\varphi$ can be determined with a precision of
\be
\langle (\hat{\varphi}-\varphi)^2\rangle \approx \frac{\pi^2}{N^2} \label{eq:BW}
\ee
 with a single shot measurement and a $N$-qubit state. The generator $\hat{G}_{BW}$ has a linear spectrum with $L=N+1$ different eigenvalues and a spectral range of $\Delta=N$ leading to the spectral decomposition
\be
\hat{G}_\text{BW}=\sum\limits_{\mu=1}^{N+1}(\mu+c)\ket{\mu}\bra{\mu}
\ee
with $c=-N/2-1$ being an irrelevant constant which we will neglect from now on.

Here, the number of levels $L$ and the spectral range $\Delta$ are proportional to the number of qubits $N$. However, this is not necessary the case for global fields $B(r)$ with arbitrary spatial dependence $f(r)$ and different positioning of the local sensors. Therefore, we investigate in the following how $L$ and $\Delta$ influence the optimal precision separately. Thus, we generalized the results of Berry and Wiseman \cite{Berry2000} to frequency estimation and generators with rescaled linear spectrum
\begin{eqnarray}
\hat{G}_0&=&\frac{\Delta}{L-1}\sum\limits_{\mu=1}^{L}\mu\ket{\mu}\bra{\mu}\\
&=&\frac{\Delta}{L-1}\hat{G}_\text{BW}
\end{eqnarray}
where $L$ and $\Delta$ can be varied independently.

The time evolution determined by $\exp[-i\omega t \hat{G}_0]$ is equivalent to $\exp[-i\varphi\hat{G}_{BW}]$ with $\varphi=(\omega t \Delta)/(L-1)$.
We assume that the frequency $\omega$ is equally distributed between $0\leq \omega < W_0$. Letting the system evolve for
\be
t_1=\frac{2\pi}{W_0}\frac{L-1}{\Delta} \label{eq:time}
\ee
leads to an equal distribution of $\varphi$ with $0\leq \varphi < 2\pi$.
As a consequence, we can achieve a precision of
\be
\langle(\hat{\omega}-\omega)^2\rangle =\frac{\langle(\hat{\varphi}-\varphi)^2\rangle}{|\partial_\omega \varphi|^2}\approx \frac{ W_0^2}{4L^2}
\ee
determined by \eq{eq:BW}. As a result, only the number of levels $L$ is important for the precision in a single shot experiment if the measurement time can be chosen appropriately. However, the time $t_1$ to achieve this precision scales inversely with the spectral range $\Delta$. Therefore, maximizing the number of levels is only optimal when the interaction time $t_1$ can be chosen arbitrarily.  However, not only  the number of qubits is a resource, but in general also time, which we will investigate in the next section.


\subsection{Multi-shot estimation}
In this section, we investigate scenarios where the total interaction time $T$ is fixed. Here, we assume that $T$ can be split between different measurements. A basic approach would to repeat the measurement described in the previous section $\nu=T/t_1$ times without updating the prior. Since the precision scales with $ 1/\nu$, we arrive finally at
\be
\langle(\hat{\omega}-\omega)^2\rangle \sim \frac{W_0^2 t_1}{4L^2T} \sim \frac{W_0 }{2T L\Delta}.
\ee
As a result, the optimal precision scales inverse with the product of the number of levels $L$ times the spectral range $\Delta$ of the generator $\hat{G}_0$.

A better measurement scheme would update the prior $p(w)$ after each measurement and adapt the probe state, evolution time and measurement for each run.
Assuming that the frequency distribution stays flat at each round we can now define a sequence of  interaction times and widths $(t_k;W_k)$ with
\be
W_k = W_0(2L)^{-k}
\ee
and
\be
t_k = \frac{L-1}{ \Delta}\frac{2\pi}{W_{k-1}}
= \frac{2\pi(L-1)}{W_0\Delta}(2L)^{k-1}.
\ee
As a consequence, the total interaction time after $n$ measurement rounds  is given by
\be
T =\sum\limits_{k=1}^n t_k =  \frac{2\pi(L-1)}{W_0\Delta}\frac{(2L)^{n}-1)}{2L-1}
\approx \frac{\pi}{\Delta W_0}(2L)^n.
\ee
Thus, the maximal number of estimation rounds is upper bounded by
\be
(2L)^n\leq\frac{\Delta W_0}{\pi}T
\ee
for a fixed interaction time $T$. Therefore, the maximal achievable precision is upper bounded by
\be
W_T = W_n = W_0(2L)^{-n} \geq\frac{\pi }{T\Delta }
\ee

suggesting that in such an adaptive scheme only the spectral range has an effect on the
scaling of the precision with time.


\subsection{Single-shot, fixed time estimation\label{sec:single_shot}}

In the two previous subsections, we assumed that it is possible to arbitrarily choose and split the interaction time. However, often only short interaction times are available in real world estimation problems. In addition, also preparing a good probe state and performing measurements needs time which exceeds in some cases the actual interaction time.
As a consequence, there exist many scenarios where only a single-shot estimation with fixed interaction time is possible. In this case, our estimation problem falls into the regime of Bayesian frequency estimation \cite{Macieszczak2014,Sidhu2019}. For Gaussian prior distributions $p(\omega)$, the precision of the updated distribution after the measurement is given by \cite{Macieszczak2014}
\be
\langle(\hat{\omega}-\omega)^2\rangle=W_1^2 = W_0^2\left(1-W_0^2\cdot F(\bar{\rho},\hat{G}_0t)\right).
\ee
Here, $F$ denotes the quantum Fisher information,  $W_0$ the variance of the prior and $\bar{\rho}$ the prior weighted density operator  with matrix elements
\begin{eqnarray}
\bar{\rho}_{n,m}&=& \int \text{d}\omega \;c_nc_m^\ast \exp\left[-\I \omega t (n-m)\right] p(\omega)\\
&=& c_nc_m^\ast   \exp\left[-\frac{t^2W_0^2\Delta^2}{2(L-1)^2}(n-m)^2\right],\label{eq:rho}
\end{eqnarray}
given in the eigenbasis of $\hat{G}_0$.
The optimization of the probe state is in general non-trivial and often only possible with numerical methods and iterative algorithms \cite{Demkowicz2011,Macieszczak2014}. However, we can adapt some of the results of \cite{Macieszczak2014} by rescaling the dimensionless time parameter
\be
\tau= tW_0 \rightarrow tW_0 \frac{\Delta}{L}.
\ee
For $tW_0\Delta \ll 1$, all off-diagonal terms \eq{eq:rho} survive and thus a GHZ-like probe state is optimal. In this case, the variance reduction factor is given by
\be
\frac{W_1^2}{W_0^2}=1-t^2W_0^2\Delta^2 \exp\left(-t^2W_0^2\Delta^2\right).
\ee
As a consequence, the precision is mainly determined by  the spectral range $\Delta$ alone   as long as $tW_0\Delta \ll 1$.

For $tW_0\Delta \gg 1$, only off-diagonal elements \eq{eq:rho} with $n\approx m$ survive provided  $tW_0\Delta/L$ is  of the order of unity. Then, the number of different levels is important and states similar to the Berry-Wiseman state \cite{Berry2000} are optimal.

In the intermediate regime, states with a structure interpolating between GHZ-state and the Berry-Wiseman state are optimal. Again, we can adapt results from \cite{Macieszczak2014} to our situation. A fixed number of levels $L$ in our case corresponds to a fixed number of atoms $N$ in \cite{Macieszczak2014} with $L=2N$. Thus, for a fixed $L$ and $2\leq L\leq 40$, a ratio of $0.5\leq tW_0 \Delta/(L-1)\leq 1$ is optimal as can be seen from Fig. 2 in \cite{Macieszczak2014} suggesting that $L$ and $\Delta$ should be increased simultaneously in this regime if possible. However, a larger number of levels $L$ leads always to a higher precision provided the ratio $\Delta/L$ is fixed.

For $tW_0\Delta/L>>1$, no off-diagonal elements survive and phase estimation is not possible anymore. In this case, a shorter interaction time should be chosen.


\section{Creating linear spectra with local multilevel systems in noisy environments\label{sec:spectra}}

The advantage of quantum metrology is severely limited by the influence of noise. In the worst case scenario, the advantage shrinks to a constant factor \cite{Fujiwara2008,Escher2011,Escher2012,Kolodynski2012, Sekatski2017PRX,Sekatski2017quantummetrology}. However, the quadratic improvement of quantum metrology can be maintained in certain situations by using e.g. error correction or fast control \cite{Sekatski2017PRX, Sekatski2017quantummetrology,Dur_2014,Arrad_2014,Kessler_2014,Sekatski_2016,Zhou2018,Altenburg2016,Landini2014}. In \cite{Sekatski2019}, we have described how to protect global parameter estimation from noise sources with given spatial distributions by designing appropriate probe states. These states consisted of superpositions of effective energy levels of the generator $\hat{G}_0$ within a decoherence-free subspace. With these probe states, it is possible to achieve the same precision scaling as in the noiseless case.

We mainly concentrated in \cite{Sekatski2019} on the Fisher regime and thus on probe states consisting of a superposition of only two orthogonal states. However, the maximal achievable precision in the Bayesian regime crucially depends  on the number of levels as we have discussed in the previous section.
It is in general a difficult task to find optimal probe states and precision limits in the Bayesian regime. Previous works \cite{Berry2000, Demkowicz2011, Macieszczak2014,Sidhu2019,Sekatski2017singlequbit} mainly concentrated on generators $\hat{G}_0$ with equally spaced levels. Therefore, we will also concentrate on this regime and investigate in the following methods to create effective linear spectra within the decoherence-free subspace. All previous results \cite{Berry2000, Demkowicz2011, Macieszczak2014,Sidhu2019,Sekatski2017singlequbit} as well as our considerations from \Sec{sec:L_vs_D} can then be adopted to probe states based solely on this  effective spectra.

In this section, we concentrate on methods based on a fixed number of sensor at fixed positions and variable internal degrees of freedom. In the next section, we will concentrate on methods based on sensors with fixed internal degrees of freedom but with variable positioning.

\subsection{One-dimensional orthogonal subspace}

Our goal is to determine the global phase $\varphi=t\omega$ generated by $\hat{G}_0$, \eq{eq:G}, with $0\leq \omega \leq W_0$ without being disturbed by global phase noise generated by $\lbrace \hat{G}_k\rbrace$ with $1\leq k\leq K$ and different spatial distributions $f_k(r_j)$.

The spatial distribution of each generator $\hat{G}_k$ can be described by the vector $\mathbf{f}_k=(f_k(r_1),\cdots,f_k(r_j))^T$ for $0\leq k\leq K$. In the following, we investigate a situation with one signal source and $K$ noise sources which are linear independent meaning that their corresponding spatial vectors $\mathbf{f}_k$ are linear independent.

The state of our sensor network can be described by time averaged spin vectors
\be
\mathbf{s}=\langle \hat{\mathbf{Z}} \rangle_t = (\langle \hat{Z}_1\rangle_t, \cdots , \langle \hat{Z}_J\rangle_t)^T
\ee
determined by the time averaged expectation values $\langle \hat{Z}_j\rangle_t$ of the local operators $\hat{Z}_j$.
In the following, we assume that $\mathbf{s}$ describe  time averaged eigenstates (compare \cite{Sekatski2017singlequbit, Sekatski2019}). This means that a system in state $\mathbf{s}$  is at all times in an energy eigenstate, however, the eigenstate might change due to spin flips during the coherent time evolution generating the phase $\varphi=\omega t$. The coherence between two different effective spin vectors $\mathbf{s}$ and $\mathbf{r}$ is preserved if \cite{Sekatski2019}
\be
\mathbf{f}_k^T(\mathbf{s}-\mathbf{r})=0 \quad  1\leq k\leq K
\ee
and the effective signal strength is given by
$
\mathbf{f}_\perp^T(\mathbf{s}-\mathbf{r}).
$
Here, $\mathbf{f}_\perp$ denotes the component of the signal $\mathbf{f}_0$ which is orthogonal to all noise vectors $\mathbf{f}_k$.

The subspace orthogonal to $\text{span}\lbrace \mathbf{f}_k\rbrace$ with $1\leq k\leq K$ is one dimensional if our sensor network consist of sensors with only $J=K+1$ different positions.
In this case, the optimal probe state consists of a superposition of spin states $\mathbf{s}$ parallel to $\mathbf{f}_\perp$ (compare with \cite{Sekatski2019}). To create non-integer multiples of $1/2$, we use intermediate spin flips such that the time average of the spin is given by
$
 \mathbf{s}=\langle \hat{\mathbf{Z}} \rangle_t \parallel \mathbf{f}_\perp.
$

In the following, we investigate methods to create an effective linear spectra described by $\mathbf{s}$ within the decoherence-free subspace. We assume that the position of all sensors are fixed and that each sensor consists of a quantum system with $n_j$ linear spaced energy levels with energies $\lbrace E_m=m\rbrace$ and $-n_j/2 \leq m \leq n_j/2$.
One possibility to create superpositions of effectively linear spaced levels is to use equivalent local sensors with $n_j=n$ energy levels and create the superposition state
\be
\ket{\psi}=\sum\limits_{m=-n/2}^{n/2} \bigotimes_{j=1}^J \ket{\text{sign}(f_\perp^j)m}_j.
\ee
For each local system $j$, we time the local spin flips such that the average energy of the level $m=+1/2$ is given by
\be
\langle 1/2 \rangle _j=\frac{|f_\perp^j|}{2f_\perp^\text{max}}
\ee
where $f_\perp^j$ denotes the component $j$ of $\mathbf{f}_\perp$ and $f_\perp^\text{max}$ the maximal component of $\mathbf{f}_\perp$.
As a consequence, each energy level $m$ is mapped to $m |f_\perp^j|/f_\perp^\text{max}$ and we arrive at the effective probe state
\be
\ket{\psi_\text{eff}}=\sum\limits_{m=-n/2}^{n/2} \bigotimes_{j=1}^J \ket{m \frac{ f_\perp^j}{f_\perp^\text{max}}}_j = \sum\limits_{m=-n/2}^{n/2} \ket{m \frac{\mathbf{f}_\perp}{f_\perp^\text{max}}}\label{eq:probestate}
\ee
which consist of a superposition of $n$ states with  effective spins $\mathbf{s}_m \parallel \mathbf{f}_\perp$. Thus, they all ly in the same decoherence-free subspace.

However, a superposition state with equal level spacing and spectral range can be also achieved with less resources if $n\cdot |f_\perp^j|/f_\perp^\text{max}\leq (n-1)$ for some $j$. In this case, we use systems of dimension $\lceil n|f_\perp^j|/f_\perp^\text{max}\rceil$ for each local system. However, at least one system still has dimension $n$. We use this system as control system for controlled spin flips on all the other systems to create the effective superposition state given in \eq{eq:probestate}. In general, it is also possible to create superpositions of $n$ levels with only single qubits for each sensor. In this case, an auxiliary system with $n$ levels which is insensitive to all fields (source and noise) is necessary to control the necessary spin flips (see \cite{Sekatski2017singlequbit}).

\subsection{Multi-dimensional orthogonal subspace}

\begin{figure}
\begin{center}
\includegraphics[width=0.5\textwidth]{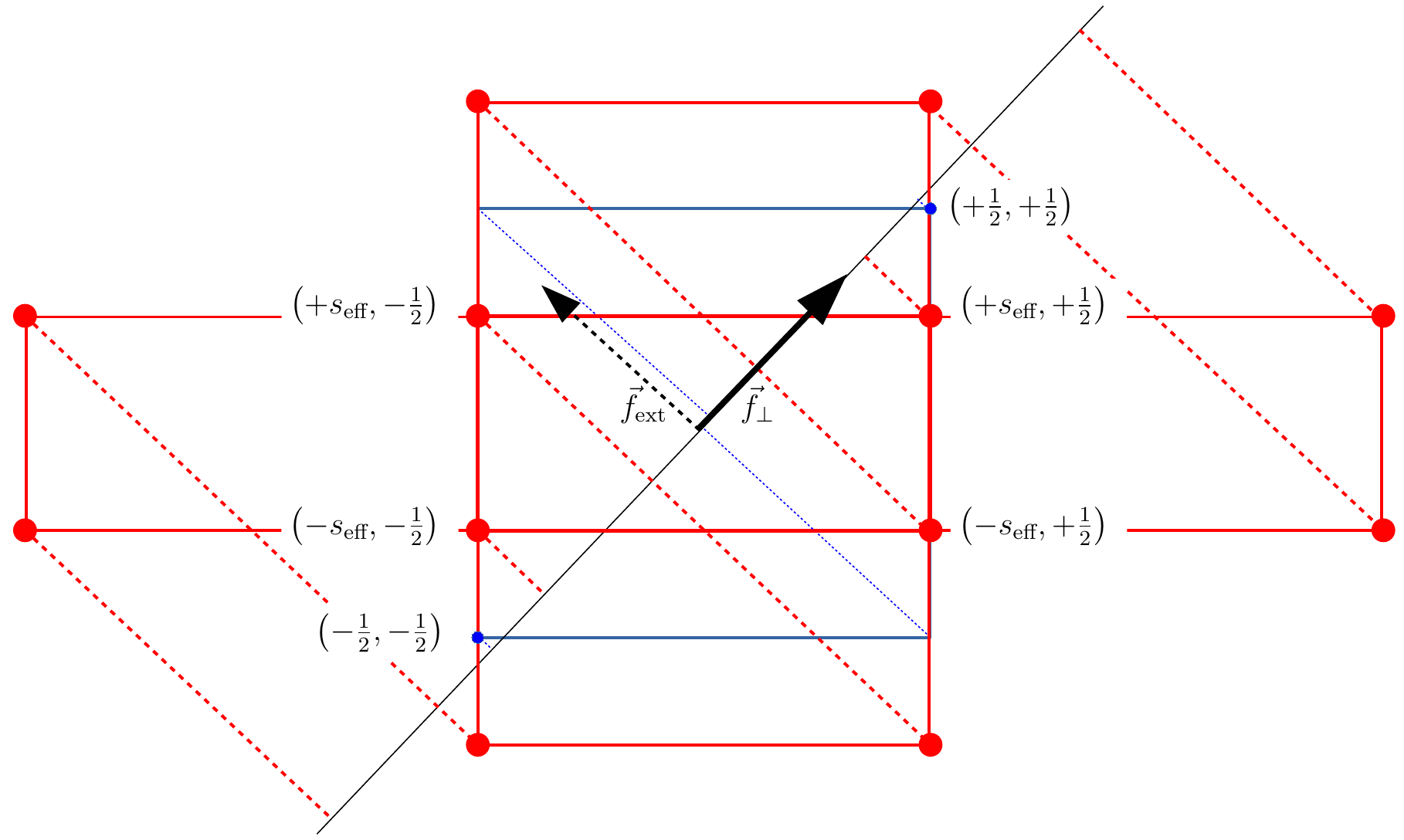}
\end{center}
\caption{Visualization of the states of a two qubit system and their projection onto the perpendicular signal component $\mathbf{f}_\perp$ (black arrow). The projections of the four spins states $(\pm s_\text{eff},\pm 1/2)$ (red dots) lead to an equally spaced level structure optimal four quantum metrology in the Bayesian regime. The superposition of the two states $(1/2,1/2)$ and $(-1/2,-1/2)$ are optimal for quantum metrology in the Fisher regime. However, their projection together with the projection of $(1/2,-1/2)$ and $(-1/2,1/2)$ do not lead to equally spaced levels.    }
\label{fig:multilevel}
\end{figure}

 There are automatically more levels available in the decoherence-free subspace if  the space orthogonal to $\text{span}\lbrace \mathbf{f}_k\rbrace$ with $1\leq k\leq K$ is multi-dimensional. This is possible if there exist more than $J>K+1$ different sensor positions. For example, in the appendix of \cite{Sekatski2019}, an example of a three qubit system is discussed, where all states of the form $\ket{\pm 1/2,\pm1/2,s_3}$ with arbitrary but fixed $s_3$ lie within the same decoherence-free subspace. In this case, the states $\ket{\mathbf{s}}$ used for the superposition probe state must not necessarily be parallel to $\mathbf{f}_\perp$. However, the projection of these states onto $\mathbf{f}_\perp$ are not necessary equidistant, see for example the projection of the states $\ket{\pm 1/2,\pm 1/2,s_3}$ onto $\mathbf{f}_\perp$ as depicted in \fig{fig:multilevel} (the state $s_3$ of the third qubit was neglected in this figure to simplify the presentation). Again, we can us dynamical controlled spin-flips to solve this problem. In this case, we decrease the spin of one of the systems from $1/2$ to $s_\text{eff}$ in such a way that
\be
\left(\begin{array}{c}s_\text{eff}\\1/2\end{array}\right) \cdot \mathbf{f}_\perp = 3 \left(\begin{array}{c}-s_\text{eff}\\1/2\end{array}\right) \cdot \mathbf{f}_\perp.
\ee
In this way, the projection of the states $\ket{\pm s_\text{eff}, \pm 1/2}$ lead to linear spaced projections onto $\mathbf{f}_\perp$. To obtain more levels, it is enough to increase the number of linear spaced energy levels  of one of the systems. In our example, we can get $2n$ ``equally'' spaced states within the decoherence-free subspace. This method can be generalized to higher dimensional decoherence-free subspaces.

As a result, we can generate linear multi-level spectra within the decoherence-free subspace for Bayesian parameter estimation by using well timed spin-flips. The maximal achievable spectral range $\Delta$ as well as the number of levels $L$ scale with the number of local energy levels $n$ similar to the noiseless case. In addition, the dimension $d=J-K$ of the space orthogonal to all noise sources can be also used to generate effective linear spectra by using again well timed spin-flips.

\subsection{Arbitrary effective spectrum}

We have concentrated on generating effective linear spectra w.r.t. the signal generating Hamiltonian so far, as such spectra are typically used in a Bayesian estimation scenario with flat prior distribution. However, there are many other metrological scenarios as well, and the optimal states and optimal effective energy spectra vary from case to case. Hence we also discuss a general method to obtain arbitrary effective spectra within a DFS using dynamical control. Once one has constructed a two-dimensional DFS with eigenstates $|v^+\rangle,|v^-\rangle$ and eigenvalues $\lambda^\pm=\pm\Delta/2$ where $\Delta$ is the spectral range, one can obtain a multi-dimensional DFS with degenerate eigenvalues by simply placing more sensors (or a higher dimensional system) at each sensor position. Similarly, adding auxiliary systems that are not taking part in the sensing process have a similar effect. We assume in the following that each eigenstate is $k$-fold degenerate, $|v_k^\pm\rangle = |v^\pm\rangle|k\rangle$, with eigenvalues $\lambda_k^\pm = \lambda^\pm$. By performing a controlled-switch between eigenstates $|v^+_k\rangle$ and $|v_k^-\rangle$ at appropriated times, one can generate effective eigenvalues $\tilde\lambda_k^+, \tilde\lambda_k^-$ with arbitrary values $0 \leq \tilde\lambda_k^+ \leq \Delta/2$ and $\lambda_k^-=-\lambda_k^+$. This allows one to produce an arbitrary symmetric spectrum. An arbitrary asymmetric spectrum can be obtained by mixing each of the eigenstates separately with an effective zero-energy state. Notice that effective zero energy levels can be generated by using two other auxiliary levels. An alternative is to use just the positive part of the spectrum, which however results in a decrease of the spectral range by a factor of $1/2$.

A similar method works to modify a given linearly spaced spectrum $\{\pm \lambda_k\}$. By adding degeneries (e.g. using auxiliary states or levels that do not take part in the sensing process), one can either mix pairs of levels $\pm \lambda_k$ with effective zero-energy states and move energy down for the two levels. Or one can also mix two different energy levels $\lambda_{k_1},\lambda_{k_2}$, which results in moving one energy up and the other down (but also changes the spectral range eventually).


\section{Creating linear spectra within the decoherence-free subspace by varying the spatial distributions \label{sec:position}}	

\begin{table}
\begin{tabular}{|l|c|c|}
\hline
spatial distribution & $\Delta$ & $L$ \\ \hline
2-point & $N$ & $N/2$ \\ \hline
linear & $\approx N/4$ & $\approx N^2/4$ \\ \hline
exponential & $\approx 2$ & $2^{N/2}$  \\ \hline
\end{tabular}
\caption{Summary of spectral range $\Delta$ and number of levels $L$  for gradient estimation with $N$ qubits for different spatial distribution within the subspace protected from global phase noise. }
\label{tab:gradient}
\end{table}

In the previous section, we investigated how to create effectively linear spaced levels assuming a fixed number of sensors at fixed positions. Here, we increased the number of level $L$ as well as the spectral range $\Delta$  simultaneously by increasing the internal degrees of freedom of the local sensors.

However, the number of available levels $L$ and the maximal achievable spectral range $\Delta$ are strongly influenced by the positioning of the different sensors. Thus, we can increase $L$ or $\Delta$ just by varying the spatial distribution of our sensors without using additional resources such as additional qubits to increase the internal degree of freedom of the sensors. In general, increasing one  will lead to a degrease of the other. Thus, the trade-off between $L$ and $\Delta$ need to be  carefully balanced depending on the actual situation as discussed in \Sec{sec:L_vs_D}.

To be fully flexible, we present here different constructions to achieve states with up to exponentially many effective energy levels, at the prize of a (linearly) reduced spectral range. In contrast to \Sec{sec:spectra}, we now assume that each sensor is described by a single qubit. Our results can be generalized to sensor with more internal degrees of freedom by combining the methods from this section and \Sec{sec:spectra}.

We start by  concentrating on gradient estimation with the generator
     \be
     \hat{G}_0=\sum\limits_j r_j \hat{Z_j},
     \ee
with  normalized positions $-1/2\leq r_j\leq 1/2$. Our goal is to determine the global phase $\varphi=t\omega$ generated by $\hat{G}_0$ with $0\leq \omega < W_0$ without being disturbed by global phase noise generated by
\be
\hat{G}_1 = \sum\limits_j \hat{Z_j}.
\ee
In the Fisher scenario, it is optimal to place $N/2$ qubits at $r_j=\pm 1/2$, respectively, because we achieve in this way the maximal possible spectral range of $\Delta= N$ \cite{Altenburg2017}. However, we obtain only $L=N/2$ different eigenvalues for $\hat{G}_0$. In the following, we discuss different spatial arrangements of our sensors to generate linear spectra with different combinations of $L$ and $\Delta$ which can help to optimize global parameter estimation in the Bayesian regime.

\subsection{Linear spacing}
For simplicity, we assume that the number of qubits $N$ is even. In this case, positioning $N$ sensors with linear spacing leads to
\be
r_{\pm j} = \pm \frac{j-1/2}{N-1} \quad , \quad 1\leq j \leq N/2.
\ee
The maximal eigenvalue $\Gamma_\text{max}$ is achieved if all spins with positive $r_j$ pointing up and all others down, leading to
\be
\Gamma_\text{max}= 2\cdot \frac{1}{2}\sum\limits_{j=1}^{N/2} \frac{j-1/2}{N-1}.
\ee		
In a similar way, we find $\Gamma_\text{min}= -\Gamma_\text{max}$.	
Thus the spectral range is given by
\be
\Delta = \Gamma_\text{max} - \Gamma_\text{min} = \frac{N^2}{4(N-1)}\approx \frac{N}{4}.
\ee		
Similar considerations for $N$ odd leading to the same scaling of $\Delta\approx N/4$. The smallest energy change is achieved if the spins situated at $r_{\pm 1}$ are changed. Both of these spins need to be switch simultaneously, to stay in the protected subspace from global field noise. This leads to a minimal energy change of $\delta = 1/(N-1)$ and as a result to a maximal number of energy levels of $L=\Delta/\delta\approx N^2/4$. As a result, increasing the number of qubits leads to a similar scaling of the spectral range as in the Fisher regime while we get a quadratic improvement in the number of levels.

\subsection{Exponential spacing}
The maximal number of levels with equidistant spacing $L=2^{N/2}$ is achieved if the particles are placed at
\be
r_{\pm j}=\pm \frac{1}{2} \frac{1}{2^{j-1}}\quad ,\quad 1\leq j \leq N/2
\ee
where we again took into account that only states within the protected subspace  are interesting. In this case, the maximal and minimal eigenvalues are given by
\be
\Gamma_\text{max/min}=\pm \frac{2}{4}\sum\limits_{j=1}^{N/2} \frac{1}{2^{j-1}} = \pm\left(1-\frac{1}{2^{N/2}}\right)
\ee
leading to a spectral range of $\Delta \approx 2 $ for large $N$. As a consequence, the achievable spectral range $\Delta$ is limited in this case and cannot be enhanced above a certain threshold by increasing the number of qubits. However, the precision depends only on $L$ for single shot estimation if the interaction time t is large enough such that $t W_0 \Delta \gg 1$ (see \Sec{sec:single_shot}). In this case, it is not necessary to increase $\Delta$ and we profit from the exponential scaling of $L$.

\subsection{Arbitrary functions}

The above conducted considerations can be generalized to arbitrary generators
\be
\hat{G}_0=\sum\limits_j f(r_j) \hat{Z}_j.
\ee
Again, our goal is to construct probe states, which are insensitive to global phase noise generated by $\hat{G}_1=\sum_j \hat{Z}_j$. However, now we do not demand that the positioning of the sensors itself is linear or exponential, but the resulting field strengths $f(r_j)$ when hopping from one sensor to another. In this way, we can generate similar level structures as in the case of gradient estimation.

To achieve linear spacing,  the sensors need to be placed at positions $r_j$ such that
\be
f(r_j)=f_j=a \frac{j-1/2}{N-1}+b.
\ee
To construct a probe state which is insensitive to global field noise, we determine the component $\mathbf{f}_\perp$ of $\mathbf{f}$ which is orthogonal to the vector $\mathbf{f}_0=(1,\cdots,1)^T$ describing global field noise \cite{Sekatski2019}. This component is given by
\begin{eqnarray}
f_\perp^j= f(r_j)-\frac{\sum\limits_{j=1 }^N f(r_j)}{N}=a \frac{j-1/2-N/2}{N-1}
\end{eqnarray}
and is antisymmetric such that $f_\perp^j =- f_\perp^{N+1-j}$. The largest eigenvalue of $\hat{G}_0$ is achieved if all spins for $1\leq j \leq N/2$ pointing down and all other up. The state with the smallest eigenvalue is obtained by flipping all spins leading to a spectral range of
\be
\Delta = 2\cdot2\cdot \frac{1}{2} \sum\limits_{j=N/2+1}^N a \frac{j-1/2-N/2}{N-1}= a \frac{N^2}{4(N-1)}.
\ee
The smallest changes within the protected subspace is achieved if the two middle spins ($j=N/2$ and $j=N/2+1$) are switched leading to a level spacing of $\delta=a/(N-1)$ and in total $L=N^2/4$ levels similar to the case of gradient estimation with linear positioned sensors.

For arbitrary function $f(r_j)$ it is also possible to create $L=2^{N/2}$ equidistant levels within a decoherence-free subspace with  $N$ qubits as we will demonstrate in the following. So far, we have always used the fact that two qubits with opposite spin form a 2-dimensional decoherence-free  subspace. That is, always two qubits form one logical qubit with
\begin{eqnarray}
\ket{+}_{L_j}&=&\ket{+}_j\ket{-}_{-j} \\
\ket{-}_{L_j}&=&\ket{-}_j\ket{+}_{-j}
\end{eqnarray}
leading to
\begin{eqnarray}
\hat{G}_1\ket{\pm}_{L_j}&=&0\\
\hat{G}_0\ket{\pm}_{L_j}&=&\pm(f(r_j)-f(r_{-j}))\ket{\pm}_{L_j}.
\end{eqnarray}
Here,  $\ket{\pm}$ denotes a spin-eigenstate with the spin pointing up or down, respectively.
To create $2^{N/2}$ levels within the decoherence-free subspace, we need $N/2$ independent pairs $(j,-j)$ of sensors with
\be
f(r_j)-f(r_{-j})= \frac{a}{2^{j-1}} \quad, \quad 1\leq j\leq N/2.
\ee
The maximal and minimal eigenvalues are then given by
\be
\Gamma_\text{max/min}=\pm \frac{1}{2}\sum\limits_{j=1}^{N/2} \frac{a}{2^{j-1}}=\pm a(1-\frac{1}{2^{N/2}})
\ee
leading to a spectral range of $\Delta \approx 2a$ for large $N$. Finding the positions $r_{\pm j}$ is straightforward if $f$ is continuous and an inverse function $f^{-1}$ is known ($f^{-1}$ need not necessarily be unambitious). The pairs $(j,-j)$ can be freely chosen since the function $f_0(r)$
describing the global noise is constant. The optimal strategy is to choose $r_1$ such that $f(r_1)$ is equal to the maximum of $f(r)$ within the area of allowed sensor positions and $r_{-1}$ denotes the position of the minimum. All other sensor positions are then consecutively defined via
\be
f(r_{\pm j})= \frac{1}{2}\left(f_\text{max}+f_\text{min}\pm \frac{f_\text{max}-f_\text{min}}{2^{j-1}}\right).
\ee
As a result, we can achieve an exponential scaling for the precision of a global field with arbitrary spatial dependence $f(r)$ in the presence of global phase noise in the Bayesian regime.


\section{Conclusion\label{sec:conclusion}}

In this paper, we first discussed the influence of the spectral range $\Delta$ and the number of levels $L$, covered by a probe state and defined by a generator $\hat{G}$ of an unknown phase $\varphi=\omega t$, on the precision to estimate $\omega$. The optimal precision is solely determined by the spectral range $\Delta$ if the interaction time $t$ between the unknown field $B=\omega f_0(r)$ with strength $\omega$ and the sensor network is very small such that $t\ll 1/(W_0 \Delta)$. Here, $W_0$ determines the width of the prior of $\omega$. This is also the case if the interaction time can be split into multiple measurements with arbitrary small interaction times. In this case, the information gain achieved by probe states based on multi-level states is compensated by longer interaction times for each single measurement (see \eq{eq:time}). Thus, it is possible to either perform a few longer measurements providing more information, due to a larger $L$, or many short measurements providing each only a single bit of information, if $L=2$. However, the total amount of available information stays constant and the precision  $\langle(\hat{\omega}-\omega)^2\rangle $
 depends solely on the spectral range $\Delta$.

However, we are limited in many situations to single-shot estimation due to preparation and measurement times longer than the available interaction time. Then, the number of levels $L$ becomes more and more important as $tW_0\Delta$ grows.

As a consequence, it is optimal to invest the given resources, e.g.  number of available qubits $n$, in different ways depending on the given estimation situation. Here, we also took the influence of different noise sources with given spatial distributions $f_k(r)$ into account. To generate a maximal spectral range $\Delta$ it is optimal put as many qubits as possible at a position with maximal effective signal strength $f_\perp^\text{max}$. In this case, $\Delta$ scales linear with the number of qubits at this position. However, we need at least sensors at $J=K+1$ different positions  if $K$ linear independent noise sources are present. This reduces the number of qubits at this positions dramatically.

To create a large number of levels $L$, it is optimal to place each qubit at a different position. Depending on the spatial distribution of the sensors, we can either maximize $L$ by sacrificing $\Delta$. In this case, we can get an exponential scaling of $L$ with the number of qubits $n$. However, $\Delta$ will be limited by a constant in this case. In other cases, it is optimal to increase $L$ and $\Delta$ simultaneously as discussed in \Sec{sec:single_shot}. In such situations, it is e.g. possible to achieve quadratic scaling of $L\sim n^2$ and still linear scaling of $\Delta\sim n$.

\section*{Acknowledgment}
This work was supported by the Austrian Science Fund (FWF) through project P30937-N27.


\bibliographystyle{apsrev4-1}
\bibliography{metrology}

\end{document}